\documentclass[manuscript]{acmart}
\acmConference[CHI '24]{}{May 11–16, 2024}{Honolulu, HI, USA}
\usepackage{tikz}
\usepackage{graphicx}
\usepackage{caption}

\AtBeginDocument{%
  \providecommand\BibTeX{{%
    \normalfont B\kern-0.5em{\scshape i\kern-0.25em b}\kern-0.8em\TeX}}}

\setcopyright{acmcopyright}
\copyrightyear{}
\acmYear{2024}
\acmDOI{}

\ccsdesc[500]{Computing methodologies~Multi-agent reinforcement learning}
\ccsdesc[500]{Computing methodologies~Sequential decision making}
\ccsdesc[300]{Information systems~Recommender systems}
\ccsdesc[100]{Human-centered computing~Collaborative and social computing theory, concepts and paradigms, Health
informatics}

\begin{document}

\title{Towards Effective Multidisciplinary Health and HCI Teams based on AI Framework}
\author{Mohammed Almutairi}
\affiliation{University of Notre Dame
 \city{Notre Dame}
 \country{USA}}
\email{malmutai@nd.edu}
\email{dgomezara@nd.edu}

\author{Diego Gómez-Zará}
\affiliation{University of Notre Dame
 \city{Notre Dame}
 \country{USA}}
\email{malmutai@nd.edu}
\email{dgomezara@nd.edu}

\keywords{\textit{Cross-Disciplinary Collaboration, health research, HCI research, collaborative research}}

\maketitle

\section{MOTIVATION}
As a Ph.D. student with a diverse background in both public and private sectors, I have encountered numerous challenges in cross-disciplinary and multi-stakeholder team projects. My research on developing team compositions that involve multidisciplinary members from fields including education, academia, and health \cite{almutairi2024aienhanced}. Along with my advisor, we are focused on exploring how HCI can help individuals assemble more effective teams. This effort involves developing socio-technical systems that guide and inform individuals of the potential teams that they can assemble. We employ state-of-the-art algorithms that prioritize inclusion among team members from diverse areas of expertise and familiarity between the team members \cite{almautaiir2024optimizing}. Our goal for attending this workshop is to engage in meaningful dialogues with scholars and researchers, leveraging these interactions to refine our approach to building an AI-driven team composition system to foster effective, interdisciplinary collaboration in health-focused HCI research.

\section{Introduction}

The collaboration between healthcare and HCI has led to significant advancements in medicine and technology, such as in areas of medical consultations, disease diagnosis, and research and treatment strategies ~\cite{sallam2023chatgpt}. Despite these successes, a primary challenge remains:  from acute and chronic conditions to mental health, using diverse technological solutions \cite{agapie2022using}. 
Several studies highlight key challenges in multidisciplinary collaboration in health and HCI research, such as differing goals, research methodologies, publication practices, funding strategies, organizational support, and academic and career recognition ~\cite{kilcullen2022insights, maurer2022forced}. Given the challenges highlighted in recent studies, we propose an innovative AI algorithm designed to utilize both contextual team information and individual preference to mitigate factors that often hinder multidisciplinary collaborations in healthcare and HCI by aligning team composition with the specific requirements of a research project, providing a user-centric tool to facilitate effective team assembly.

\section{Challenges in cross-disciplinary team formation}
Forming multidisciplinary teams in the fields of health and HCI presents significant challenges, primarily from the need to integrate diverse expertise and perspectives. Historical models like Tuckman's four-stage development theory \cite{tuckman1965developmental} emphasize the evolution of interpersonal relationships. Hackman and Katz \cite{hackman2010group} examine the individual attributes post-team formation. However, empirical applications demonstrating how these individual attributes inform pre-formation stages in health-focused HCI research remain limited. The complexity of forming such research teams requires integrating specialists from diverse fields, each qualified in distinct methodologies and holding unique perspectives. Aligning these varied attributes during the initial team formation stage is crucial to match the individual preferences and specialties of team members. However, Formed teams solely based on members' preferences may lead to a homogeneous perspective, potentially overlooking cultural nuances crucial in patient-centered healthcare technology. This situation exemplifies the potential pitfalls of the similarity-attraction theory \cite{montoya2013meta} and social categorization theory \cite{bergami2000self},  where the tendency to work with similar individuals could result in a lack the diversity of thought and experience necessary for inclusive solutions in multidisciplinary researches. A clear implication of such theories is the prevalence of failed IT-health projects, which are often explained by the design-reality gaps, missing focus, skill, or not addressing the root issues \cite{anthopoulos2016government, al2009taxonomy}.

Acknowledging the critical role of team diversity highlights just one aspect of the multidisciplinary HCI-health team challenges. Equally important is the organization within which these teams operate. Organizational structures significantly influence teams, which may hinder or promote collaboration through work environment~\cite{34}, adequate resources~\cite{50}, communication channels~\cite{55}, organizational culture~\cite{46}, and leadership support~\cite{64}. Multidisciplinary HCI-health researchers come from different types of organizations, (e.g., research versus non-research organizations), each with its own set of challenges. Some researchers navigate hierarchical organizational structures that, while possibly reducing conflict through clear, top-down communication, may also limit collaborative innovation. Other researchers benefit from flatter organizational structures that encourage open communication and collaborative innovation ~\cite{65}.

However, given the fact that multidisciplinary research is often considered the area for innovative research that is the key to future economic growth, the balance between creativity and adherence to formal work norms presents a significant challenge for HCI and health researchers, potentially impacting their future goals, evaluations, achievements, and reward benefits \cite{kilcullen2022insights}. A study further illustrates this issue, demonstrating that organizations and funding agencies often set divergent objectives for HCI researchers, thereby altering focus towards project completion at the expense of innovation \cite{agapie2022using}. Such an environment may prioritize certain attributes (e.g., project deliverables,  timeline adherence) over others (e.g., exploratory research, innovation), potentially influencing the direction and outcomes of HCI-health research. Despite some organizations initiating cross-disciplinary centers to mitigate organizational challenges and promote research integration \cite{stokols2010cross}, the issue of favoring a limited set of attributes persists. We argue that the primary challenge in multidisciplinary collaboration in health-focused HCI is finding the optimal team. The concept of optimal teams is well-studied in the realm of multidisciplinary team formation problems such as the aerospace industry ~\cite{asatourian2021multidisciplinary} and engineering industry \cite{sedaghat2018factors}. We define an optimal team in HCI and Health as one that achieves an optimal balance between the desired individual attributes of team members and success factors, such that no objective is dominated by another, aiming to create a team that is well-connected and formed from diverse areas of expertise.

\section{Proposed Solution: AI-Driven Team Formation System \cite{almautaiir2024optimizing}}

Our proposed solution is designed to address the challenges of assembling multidisciplinary teams by utilizing team members' information to recommend the optimal team. We formulate the multidisciplinary team formation problem as a sequential decision problem, where users select teams from a pool of optimized, non-dominated team compositions. After several trials of user/team interactions, the system recommends an optimal team based on contextual information about the teams and utilizes user feedback to enhance the overall performance of team compositions. Our contribution to solving the challenge of multidisciplinary team formation is twofold. Firstly, we leverage a state-of-the-art optimization algorithm \cite{deb2002fast,jamieson2014lil} to establish predefined team compositions, effectively preventing biases (e.g., favoring expert members) during team assembly to ensure that the formation process is guided by objective criteria. Secondly, we implement a user-centric approach during the team formation process by integrating user preferences. Our approach ensures that more engaged and satisfied multidisciplinary teams collaborate.

\begin{figure}[ht]
\centering
\includegraphics[width=0.8\textwidth, height=0.3\textheight]{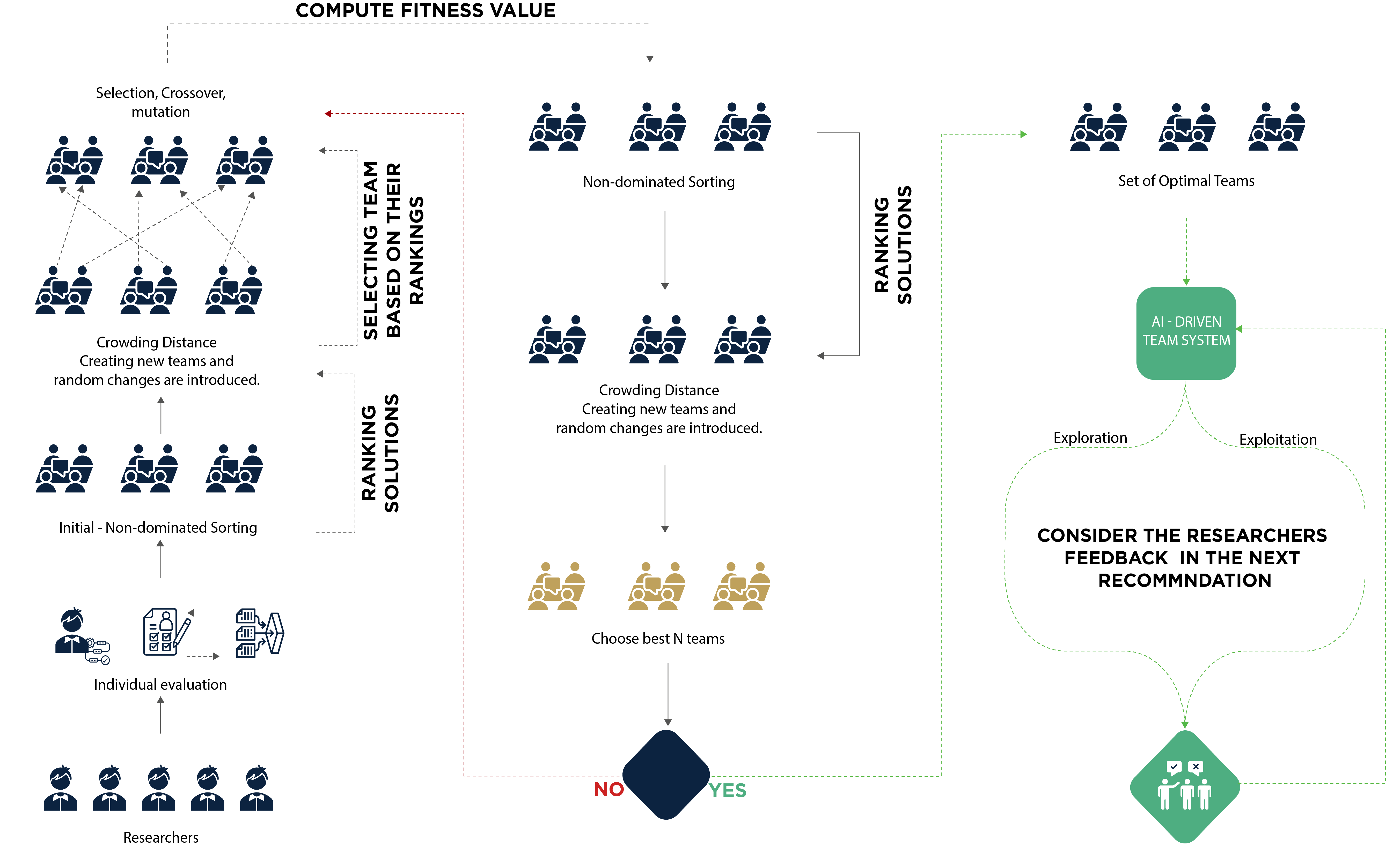} 
\caption{The figure presents an AI framework for forming multidisciplinary teams, structured in two main stages. The initial stage starts with an individual assessment of all attributes, culminating in the formation of an initial non-dominated team. Subsequently, this stage incorporates the application of essential selection, crossover, and mutation processes to cultivate a set of optimal teams. The subsequent stage involves presenting this refined set of optimal teams to the user, initiating an iterative process that continues until the selection of the desired optimal team. Throughout this selection process, the system adaptively learns the user's preferences, swiftly adjusting to highlight more preferred teams and eliminate those less desired.}
\end{figure}

\bibliographystyle{ACM-Reference-Format}
\bibliography{Citations}

\end{document}